\documentclass[a4paper,11pt]{article}
\usepackage{pos}
\usepackage{graphicx,graphics}
\usepackage{dcolumn}
\usepackage{amsmath,amsfonts}
\usepackage{latexsym,verbatim}
\usepackage{bm}
\usepackage[toc,page]{appendix}
\usepackage{color}
\usepackage[normalem]{ulem}
\usepackage{tabularx}
\graphicspath{fig}

\def\be{\begin{eqnarray}}
\def\ee{\end{eqnarray}}

\def\<{\left\langle}
\def\>{\right\rangle}

\newcommand{\beq}{\begin{equation}}
\newcommand{\eeq}{\end{equation}}

\begin{document}
\title{Electromagnetic conductivity of quark-gluon plasma at non-zero baryon density}

\author[a]{N.\,Astrakhantsev}
\author[b]{V.\,V.~Braguta}
\author[c,d]{M.\,Cardinali}
\author[c,d]{M.\,D'Elia}
\author[c]{L.\,Maio}
\author[d]{F.\,Sanfilippo}
\author*[e]{ A.\,Trunin}
\author[f]{ A.\,Vasiliev}

\affiliation[a]{Physik-Institut, Universit{\" a}t Z{\" u}rich, Winterthurerstrasse 190, CH-8057 Z{\" u}rich, Switzerland}
\affiliation[b]{Bogoliubov Laboratory of Theoretical Physics, Joint Institute for Nuclear Research, Dubna, 141980 Russia}
\affiliation[c]{Dipartimento di Fisica dell'Universit\`a di Pisa, Largo B.~Pontecorvo 3, I-56127 Pisa, Italy}
\affiliation[d]{INFN, Sezione di Pisa, Largo B.~Pontecorvo 3, I-56127 Pisa, Italy}
\affiliation[e]{Samara National Research University, Samara, 443086 Russia}
\affiliation[f]{Moscow State University, 119991, Moscow, Russia}

\emailAdd{nikita.astrakhantsev@physik.uzh.ch}
\emailAdd{vvbraguta@theor.jinr.ru}
\emailAdd{marco.cardinali@pi.infn.it}
\emailAdd{massimo.delia@unipi.it}
\emailAdd{francesco.sanfilippo@roma3.infn.it}
\emailAdd{amtrnn@gmail.com}
\emailAdd{vasiliev.av15@physics.msu.ru}

\FullConference{The 38th International Symposium on Lattice Field Theory, LATTICE2021 26th-30th July, 2021
Zoom/Gather@Massachusetts Institute of Technology}

\abstract{
In this preprint we present our results on the study of the electromagnetic conductivity in dense quark-gluon plasma obtained within lattice simulations with $N_f = 2 + 1$ dynamical quarks. We employ stout improved rooted staggered quarks at the physical
point and the tree-level Symanzik improved gauge action.
The simulations are performed at imaginary baryon chemical potential, and the
Tikhonov regularisation method is used to extract the conductivity from
current-current correlators.
Our results indicate
an increase of QGP electromagnetic conductivity with real baryon density, and this dependence is quite strong.
}

\maketitle

\section{Introduction}

The electromagnetic conductivity is a transport coefficient which parameterises charge transport phenomenon. It is believed that charge transport plays an important role in heavy-ion collision experiments since collisions of relativistic nuclei generate strong electromagnetic field which influences   
the dynamics of the created quark-gluon plasma (QGP). The electromagnetic conductivity of QGP can be studied in heavy-ion collision experiments through the measurement of the dilepton emission rate.

It is believed that QGP in heavy-ion collision experiments has nonzero baryon density which leads to the appearance of additional fermion states. These states might play an important role in the charge transport phenomena in QGP. In view of this, it is important to study how nonzero baryon density influences the electromagnetic conductivity of QGP.

The electromagnetic conductivity of quark-gluon matter with nonzero baryon density was studied in a number of papers within different phenomenological approaches (see, for instance, \cite{Steinert:2013fza,
Srivastava:2015via, 
Tripolt:2018jre, Soloveva:2019xph} and references therein).  In addition to phenomenological studies, lots of the first-principle results on the electromagnetic conductivity were obtained within lattice simulation of QCD (see, for instance, \cite{Aarts:2014nba,
Brandt:2015aqk,
Ding:2016hua,
Astrakhantsev:2019zkr} and the review of lattice results \cite{Aarts:2020dda}). Unfortunately, all lattice QCD studies were carried out at zero baryon density. 

In this preprint we are going to present the results of our lattice QCD study of the QGP electromagnetic conductivity at finite baryon density. To overcome the sign problem the simulations are carried out at imaginary baryon chemical potential and the results are analytically continued to real values of the baryon chemical potential.

\section{The lattice set-up}
In our study we consider the partition function for
${N_f=2+1}$ QCD with chemical potentials $\mu_f$ ($f = u,d,s$) coupled
to quark number operators, ${\mathcal Z}(T,\mu_u,\mu_d,\mu_s)$, in a
setup for which $\mu_u = \mu_d =\mu_B/3,\, \mu_s=0$.  The path integral formulation of
${\mathcal Z} (T,\mu_B)$, discretized via improved rooted staggered
fermions and adopting the standard exponentiated implementation of the
chemical potentials,
reads
\begin{equation}
  \mathcal{Z} = \int \mathcal{D}U e^{- \mathcal{S}_{\text{YM}}}
  \prod_{f=u,d,s}\det\left[ M_{\text{st}}^{f}(U,\mu_{f}) \right]^{1/4} ~, 
\label{partfunc}
\end{equation}
where
\begin{equation}
  \mathcal{S}_{\text{YM}} = -\frac{\beta}{3} \sum_{i,\mu\neq\nu}
  \left( \frac{5}{6}W_{i;\mu\nu}^{1\times1}
    - \frac{1}{12}W_{i;\mu\nu}^{1\times2} \right)
\end{equation}
is the tree-level Symanzik improved action
($W_{i;\mu\nu}^{n\times m}$ stands for the trace of
the $n\times m$ rectangular parallel transport in the $\mu$-$\nu$
plane and starting from site $i$), and the staggered fermion matrix is
defined as
\begin{eqnarray}
  M_{\text{st}}^{f}(U,\mu_{f})
  & = &
        am_f\delta_{i,j} +
        \sum_{\nu=1}^{4} \frac{\eta_{i;\nu}}{2}\big[
        e^{ a\mu_{f}\delta_{\nu,4}}U_{i;\nu}^{(2)}
        \delta_{i,j-\hat{\nu}}
        \nonumber\label{fermatrix}\\ 
  & - & e^{-a\mu_{f}\delta_{\nu,4}}U_{i-\hat{\nu};\nu}^{(2)\dagger}
        \delta_{i,j+\hat{\nu}}
        \big]~,
\end{eqnarray}
where $U_{i;\nu}^{(2)}$ are two-times stout-smeared links, with
isotropic smearing parameter $\rho=0.15$.

We consider two values of temperature $T=200,\,250\,$MeV; most simulations are carried out on a $12\times48^3$ lattice, with spacings $a=0.0820$\,fm and $a=0.0657$\,fm correspondingly. To check the lattice spacing dependence we also consider a $10\times48^3$ lattice with $a=0.0988$\,fm and $a=0.0788$\,fm. Notice that the introduction of nonzero baryon chemical potential leads to the sign problem. In order to overcome it we carry out lattice simulations with imaginary baryon chemical potential $\mu_I$ with the following values: $ {\mu_I} / {3 \pi T}=0.0, 0.140, 0.200, 0.245, 0.285$ for all lattices and lattice spacings under consideration. 

In Table~\ref{tab:params} we report the lattice parameters used in the simulations.
$O(100)$ decorrelated gauge configurations have been used for each simulation point. 
Bare parameters have been chosen so as to stay on a line of constant physics with physical quark masses.

\begin{table}[h!]
    \centering
    \begin{tabular}{|c|c|c|c|c|c|c|c|}
    \hline
     $a,\,\mbox{fm}$  & $L_s$ & $N_t$ & $T$,\,MeV & $m_la$ & $m_sa$ & $\mu_I/3 \pi T$ & $c(T)$ \\ \hline
     0.0988 & 48 & 10 &  200 & 0.0014 & 0.0394 & 
     \begin{tabular}{c}
       0.0,\,0.14,\,0.20, \\ 0.245,\, 0.285
     \end{tabular} & $0.0060(14)$
     \\ \hline
     0.0788  & 48 & 10 & 250 & 0.001119 & 0.031508 &     \begin{tabular}{c}
     0.0,\,0.14,\,0.20, \\ 0.245,\, 0.285
     \end{tabular} & $0.0086(14)$
     \\ \hline
     0.0820 & 48 & 12 & 200 & 0.001168 & 0.032872 &
     \begin{tabular}{c}
     0.0,\,0.14,\,0.20, \\ 0.245,\, 0.285
     \end{tabular} & $0.0076(14)$
     \\ \hline
     0.0657 & 48 & 12 & 250 & 0.000917 & 0.025810 &
     \begin{tabular}{c}
     0.0,\,0.14,\,0.20, \\ 0.245,\, 0.285
     \end{tabular} & $0.0084(10)$
     \\ \hline
    \end{tabular}
    \caption{Parameters used in the numerical simulations and the values of the coefficient $c(T)$ from formula (\ref{delta_sigma}).}
    \label{tab:params}
\end{table}

To calculate the conductivity we follow our previous work \cite{Astrakhantsev:2019zkr}.
The measurement of conductivity  consists of two parts: correlation function measurement and spectral function extraction via the Kubo formula inversion. The correlation function reads
\beq
C_{ij}(\tau) = \frac {1} {L_s^3} \langle J_i(\tau) J_j(0) \rangle,
\label{stag}
\eeq
where $\tau$ is the Euclidean time and $J_i(\tau)$ is the conserved current
\beq
\begin{split}
J_{i}(\tau)= \frac 1 4 e \sum_f q_f \sum_{\vec x} \eta_{i}(x) \bigl ( \bar \chi^f_{x} e^{ a\mu_{f}\delta_{\nu,4}} &U^{(2)}_{x,i} \chi^f_{x+i} + \\
&\bar \chi^f_{x+i} e^{-a\mu_{f}\delta_{\nu,4}} U^{(2)\dagger}_{x,i}\chi^f_{x} \bigr ),
\label{eq:current}
\end{split}
\eeq
where $x=(\tau, \vec x)$, $\eta_i(x)=(-1)^{x_1+..x_{i-1}}$, $i=1,2,3$, $\bar \chi^f_{x}, \chi^f_{x}$ are staggered fermion fields of $f=u,\,d,\,s$ flavours.

The staggered fermion correlator~(\ref{stag}) corresponds to two different operators for the even $\tau=2n \times a$ and odd $\tau=(2n+1) \times a$ slices. In the continuum limit $C_{ij}(\tau)$ reads

\beq
\label{correlator}
C^{\mbox{\footnotesize{e}},\,\mbox{\footnotesize{o}}}_{ij}(\tau)=\sum_{\vec x} \left(\langle A_i(x) A_j(0) \rangle - s^{\mbox{\footnotesize{e}},\,\mbox{\footnotesize{o}}} \langle B_i(x) B_j(0) \rangle \right),
\eeq
where $s^{\mbox{\footnotesize{e}},\,\mbox{\footnotesize{o}}} = (- 1)^{\tau}$ is the timeslice parity and 
\beq
\nonumber
A_i= e \sum_f q_f \bar \psi^f \gamma_i \psi^f, \quad B_i = e \sum_f q_f \bar \psi^f \gamma_5 \gamma_4 \gamma_i  \psi^f,
\eeq
and $\psi^f$ is Dirac spinor of the flavor $f$.  Notice that the operator $A_i$ corresponds to electromagnetic current in the continuum whereas we would like to remove the $B_i$ contribution.

Next let us recall that the current-current Euclidean correlators both for even and odd slices $C^{\mbox{\footnotesize{e}},\,\mbox{\footnotesize{o}}}_{ij}$ are related to its spectral functions $\rho^{\mbox{\footnotesize{e}},\,\mbox{\footnotesize{o}}}_{ij}(\omega)$ as
\begin{equation}
	C^{\mbox{\footnotesize{e}},\,\mbox{\footnotesize{o}}}_{ij}(\tau)=\int_0^{\infty} \frac{d\omega}{\pi} K(\tau, \omega) \rho^{\mbox{\footnotesize{e}},\,\mbox{\footnotesize{o}}}_{ij}(\omega),
	\label{eq:Kubo}
\end{equation}
where $K(\tau, \omega)=\frac{\cosh\omega(\tau - \beta/2)}{\sinh\omega\beta/2}$, $\beta=1/T$. The electromagnetic conductivity $\sigma_{ij}$ is related to the spectral densities $\rho^{\mbox{\footnotesize{e}},\,\mbox{\footnotesize{o}}}_{ij}(\omega)$ through the Kubo formulas
\begin{equation}
	\frac{\sigma_{ij}}{T} = \frac{1}{2 T} \lim\limits_{\omega \to 0} \frac 1 {\omega} \biggl ( {\rho^{\mbox{\footnotesize{e}}}_{ij}(\omega)} + {\rho^{\mbox{\footnotesize{o}}}_{ij}(\omega)} \biggr ). 
	\label{eq:limit}
\end{equation}

The contribution of the correlator $\langle B_i(\tau) B_j(0) \rangle$ to the sum $\rho^{e}_{ij}+\rho^{o}_{ij}$ cancels out and in the continuum limit the electromagnetic conductivity is reproduced. Similarly to~\cite{Aarts:2014nba,
Brandt:2015aqk,
Ding:2016hua,
Astrakhantsev:2019zkr} 
in this calculation of the correlation function (\ref{stag}) only connected diagrams are taken into account.

\section{Calculation of the  electromagnetic conductivity}

Given the correlation functions $C_{ij}^{\mbox{\footnotesize{e}},\,\mbox{\footnotesize{o}}}(\tau)$ one needs to invert the integral equation~(\ref{eq:Kubo}) and determine spectral functions $\rho_{ij}^{\mbox{\footnotesize{e}},\,\mbox{\footnotesize{o}}}(\omega)$ to find the conductivity. In this paper we going to apply model independent approach which is called Tikhonov regularization (TR)~\cite{Tikhonov:1963} method to extract the conductivity. A similar approach widely used for the reconstruction of spectral densities is Backus-Gilbert (BG) method ~\cite{Backus:1}. These methods were applied for the reconstruction of the shear and bulk viscosities  \cite{Astrakhantsev:2017nrs, Astrakhantsev:2018oue}, electromagnetic conductivity \cite{Brandt:2015aqk, Astrakhantsev:2019zkr} of QGP. Our study shows that both approaches give equivalent results for the electromagnetic conductivity of QGP at finite baryon density. For this reason to obtain  the results presented below we apply TR method.

The TR method is a non-parametric linear approach which can be used to study the spectral function. This method is aimed at the solution of the equation 
\beq
C(\tau) = \int\limits_{0}^{+\infty}\frac{d\omega}{2 \pi} \frac{\rho(\omega)}{f(\omega)} K(\tau,\omega), 
\label{eq:integral}
\eeq
where $ K(\tau,\omega) = \frac{\cosh \omega \left(\tau - \beta / 2\right)}{\sinh\omega \beta / 2} f(\omega)$, and $f(\omega)$ is an arbitrary function. 
In linear methods, one reconstructs the estimator $\bar \rho(\bar \omega)$ of the spectral function in the following form: 
\beq
\bar \rho(\bar \omega) = f(\bar \omega) \sum_i q_i(\bar \omega) C(\tau_i),
\label{barrho} 
\eeq
where $q_i(\bar\omega)$ are some functions, which exact expressions will be discussed later. Combining  Eqs.~(\ref{eq:integral}) and ~(\ref{barrho}), one gets the following relation between the estimator $\bar \rho(\bar \omega)$ and the spectral function $\rho(\omega)$:
\beq
\bar \rho(\bar \omega) = f(\bar \omega) \int\limits_0^{\infty} d \omega\, \delta (\bar \omega, \omega) \frac {\rho(\omega)} {f(\omega)}, 
\label{barf}
\eeq
where the resolution function $\delta(\bar \omega, \omega)$ is given by the formula:

\beq
\delta(\bar \omega, \omega) = \sum_i q_i(\bar \omega) K(x_i, \omega).
\eeq

If the resolution function has a sharp peak around $\bar \omega$ and normalized to $1$, according to Eq.~(\ref{barf}) the estimator $\bar \rho(\bar\omega)$ is a very good approximation to the spectral function $\rho(\omega)$. E.g., if $\delta(\omega, \bar \omega) = \delta ( \omega - \bar \omega)$ the estimator of the spectral function would exactly reproduce the spectral function $\bar \rho(\bar \omega) = \rho(\bar \omega)$.
In real calculation the resolution function has a peak of finite width of few $T$,
thus the estimator $\bar\rho(\bar\omega)$ averages the spectral function over the region of several $T$.
In particular, in our calculations at $\bar \omega = 0$ the width of the resolution function is $\sim 4\,T$. 

Now we would like to note that the TR method can reliably reconstruct $\rho(\omega = 0)$ if the resolution function $\delta(\bar \omega=0, \omega)$ is narrower than the characteristic variation scale of $\rho(\omega)$. Correlation functions of the electromagnetic currents are well described by either the ansatz combining the transport peak at small frequencies and UV contribution at large frequencies~\cite{Aarts:2014nba, Brandt:2015aqk,Ding:2016hua} or by the AdS/CFT spectral function~\cite{Ding:2016hua}.
Within the temperature interval considered in this paper the widths of $\delta(\bar \omega=0, \omega)$ are close to or smaller than the variation scale of $\rho(\omega)$ obtained in~\cite{Aarts:2014nba, Brandt:2015aqk,Ding:2016hua}. For this reason we believe that both approaches give reliable results for the conductivity extracted from such spectral functions.

Further let us discuss how one should select functions $q_i(\bar\omega)$ in Eq.~(\ref{barrho}). One of the reasonable ways is to require the minimization of the width of $\delta(\bar \omega, \omega)$. In particular, one finds the functions $q_i(\bar\omega)$ which minimize the the following functional:
\beq
\mathcal{A} = \int_0^{\infty} d \omega\, \delta^2(\bar \omega, \omega) (\omega - \bar \omega)^2.
\eeq
The minimization procedure gives the following expressions   
\begin{gather}
q_i(\bar\omega)  =  \frac { \sum_j W^{-1}_{ij} (\bar \omega) R(x_j) } { \sum_{kj} R(x_k) W^{-1}_{kj} (\bar \omega) R(x_j) }, \\
W_{ij}(\bar \omega)  =  \int\limits_{0}^{\infty} d \omega\, K(x_i,\omega) (\omega- \bar \omega)^2 K(x_j,\omega), \\
R(x_i)  =  \int\limits_0^{\infty} d \omega\, K(x_i,\omega). 
\label{eq:BG_formulas}
\end{gather}

\begin{figure}[h!]
    \centering
    \includegraphics[width=0.5\textwidth]{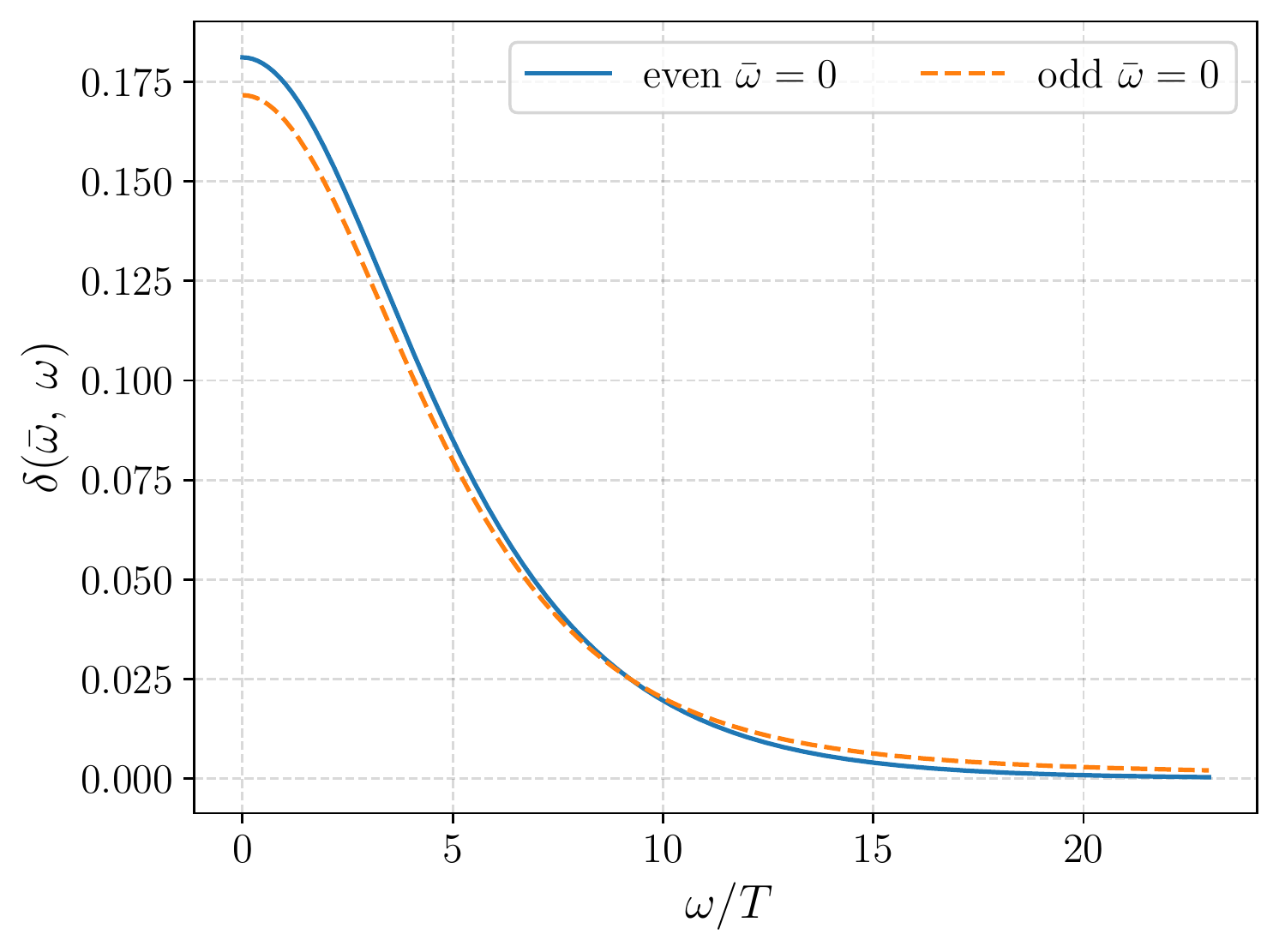}
    \caption{Typical resolution functions for the even and odd branches at $T = 200\,\mbox{MeV}$.}
    \label{fig:deltas}
\end{figure}

Notice, however, that the narrower the resolution function $\delta(\bar \omega, \omega)$ the larger number of terms in formula (\ref{barf}) with alternating sign. For this reason the method become unstable and susceptible to noise in the data. Thus, it is required to carry out regularization that should be properly adjusted. In the TR 
method one regularises the SVD decomposition of $W^{-1} = V D U^{T}$. The diagonal matrix $D = \mbox{diag}\left(\sigma_1^{-1}, \sigma_2^{-1}, \ldots, \sigma_n^{-1}\right)$ might have very large entries that represent the susceptibility of the data to noise. The regularization is done by adding the regularizer $\gamma$ to all entries as $\tilde D = \mbox{diag}\left((\sigma_1 + \gamma)^{-1}, (\sigma_2 + \gamma)^{-1}, \ldots, (\sigma_n + \gamma)^{-1}\right)$. Thus, small $\sigma_i$ will be smoothly cut-off. In Fig.~\ref{fig:deltas} we plot typical resolution functions for the TR regularization at $\bar \omega=0$.

In the reconstruction procedure one has to choose the value of the parameter $\gamma$. In order to do this we follow the approach proposed in paper \cite{Astrakhantsev:2019zkr}. Briefly, this approach can be described as follows. In the region of sufficiently small $\gamma$ (weak regularization) the method becomes unstable what manifests itself in large statistical uncertainties rising with the decrease of $\gamma$. In addition, in the same region the reconstructed conductivities and the widths of the even and odd resolution functions reach plateaus. Since the resolution function width does not decrease below some value of $\gamma$, the decrease of the $\gamma$ below this value does not make the reconstruction more accurate. On the contrary, in the region of large $\gamma$ the method is stable with small uncertainties at the cost of strong regularization. The widths of the resolution functions are growing and spectral function is averaged over a large region which is not appropriate for the conductivity reconstruction.  
In order to safely choose the region of regularization uncertainty, we restrict ourselves from below by the value of $\gamma$ where the plateaus of the reconstructed conductivities and the widths are saturated and from above by the value of $\gamma$ where both resolution functions have the width smaller than $4.7 T$. Our calculations show that these restrictions give safe estimation of the uncertainties. This results in the variation region roughly $\gamma \in (0.1, 10)$. The exact regions depend on the lattice parameters (chemical potential and temperature).

Thus, we restrict ourselves to the region $\gamma \in (0.1, 10)$  where the method is stable and the resolution is sufficiently narrow $ \sim 4T$. The uncertainties of our results were estimated as the total variation of the conductivity including statistical uncertainties as the $\gamma$ parameter is varied within the region $\gamma \in (0.1, 10)$.

Having described the TR method we proceed to the calculation of the electromagnetic conductivity which is carried out in the following steps. Firstly, we measure the lattice correlation functions $C^{\mbox{\footnotesize{e}},\,\mbox{\footnotesize{o}}}_{ij}(\tau)$ (\ref{correlator}). Then we calculate the estimators ${\bar \rho^{\mbox{\footnotesize{e}},\,\mbox{\footnotesize{o}}}(\bar \omega)}/{\bar \omega}$ at $\bar \omega =0$ within the TR approach. Finally using Eq.~(\ref{eq:limit}) we calculate the electromagnetic conductivity.

An important issue in the calculation is the ultraviolet (UV) contribution to the reconstructed conductivity.  For the conductivity the UV contribution scales as $\rho \propto \omega^2$ and our study carried out in \cite{Astrakhantsev:2019zkr} shows that the UV gives $\sim 20-30\,\%$ contribution at $\bar \omega=0$.

In principle, one could subtract the UV contribution from the correlation function (\ref{correlator}) (see \cite{Astrakhantsev:2019zkr} for details). However, this approach gives rise to a quite large uncertainty. To reduce this uncertainty we are going to apply the following approach. Instead of the correlation functions $C^{\mbox{\footnotesize{e}},\,\mbox{\footnotesize{o}}}_{\mu_I}$ we consider the difference $\Delta C^{\mbox{\footnotesize{e}},\,\mbox{\footnotesize{o}}} = C^{\mbox{\footnotesize{e}},\,\mbox{\footnotesize{o}}}_{\mu_I} - C^{\mbox{\footnotesize{e}},\,\mbox{\footnotesize{o}}}_{\mu_I=0}$. Since, for the chosen values of the lattice spacing, the UV regime starts at $\omega_0 \sim 2\,$GeV~\cite{Astrakhantsev:2019zkr}, we note that $\mu_I \ll \omega$ for all frequencies in the UV regime and baryon chemical potential. Thus, one can consider the UV spectral function independent on the imaginary baryon chemical potential and assume that the differences $\Delta C^{\mbox{\footnotesize{e}},\,\mbox{\footnotesize{o}}}$ do not contain the UV contribution. The results for $\Delta C^{\mbox{\footnotesize{e}},\,\mbox{\footnotesize{o}}}$ turn out to be more accurate since the UV--estimation uncertainty is absent in this case. The correlator $\Delta C^{\mbox{\footnotesize{e}},\,\mbox{\footnotesize{o}}}$ is related to additional conductivity due to the presence of the imaginary baryon chemical potential. In our further study we apply the TR approach to the differences $\Delta C^{\mbox{\footnotesize{e}},\,\mbox{\footnotesize{o}}}$.

\section{Results}

\begin{figure}[h!]
    \centering
    \includegraphics[width=0.45\textwidth]{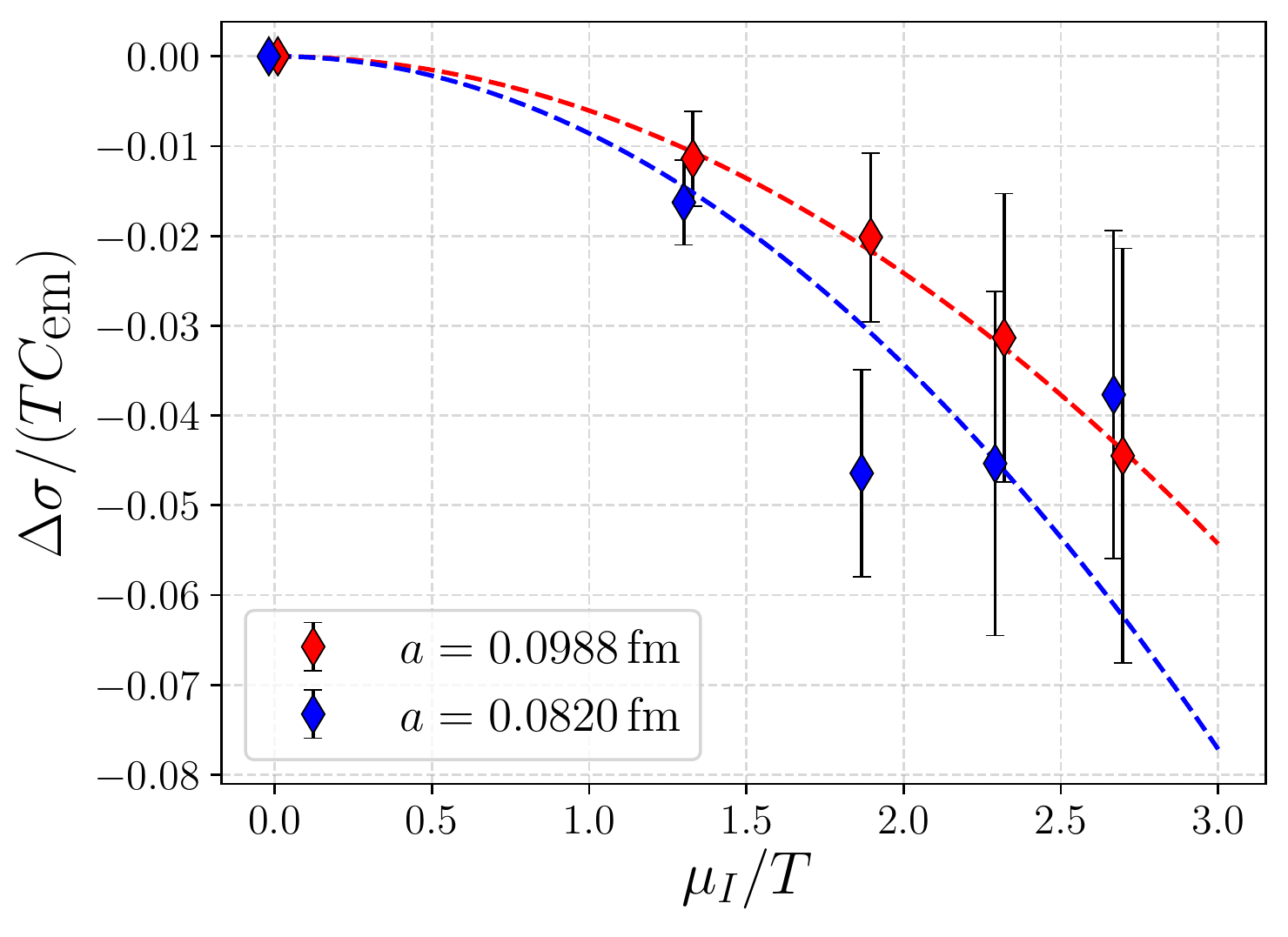}\;\includegraphics[width=0.45\textwidth]{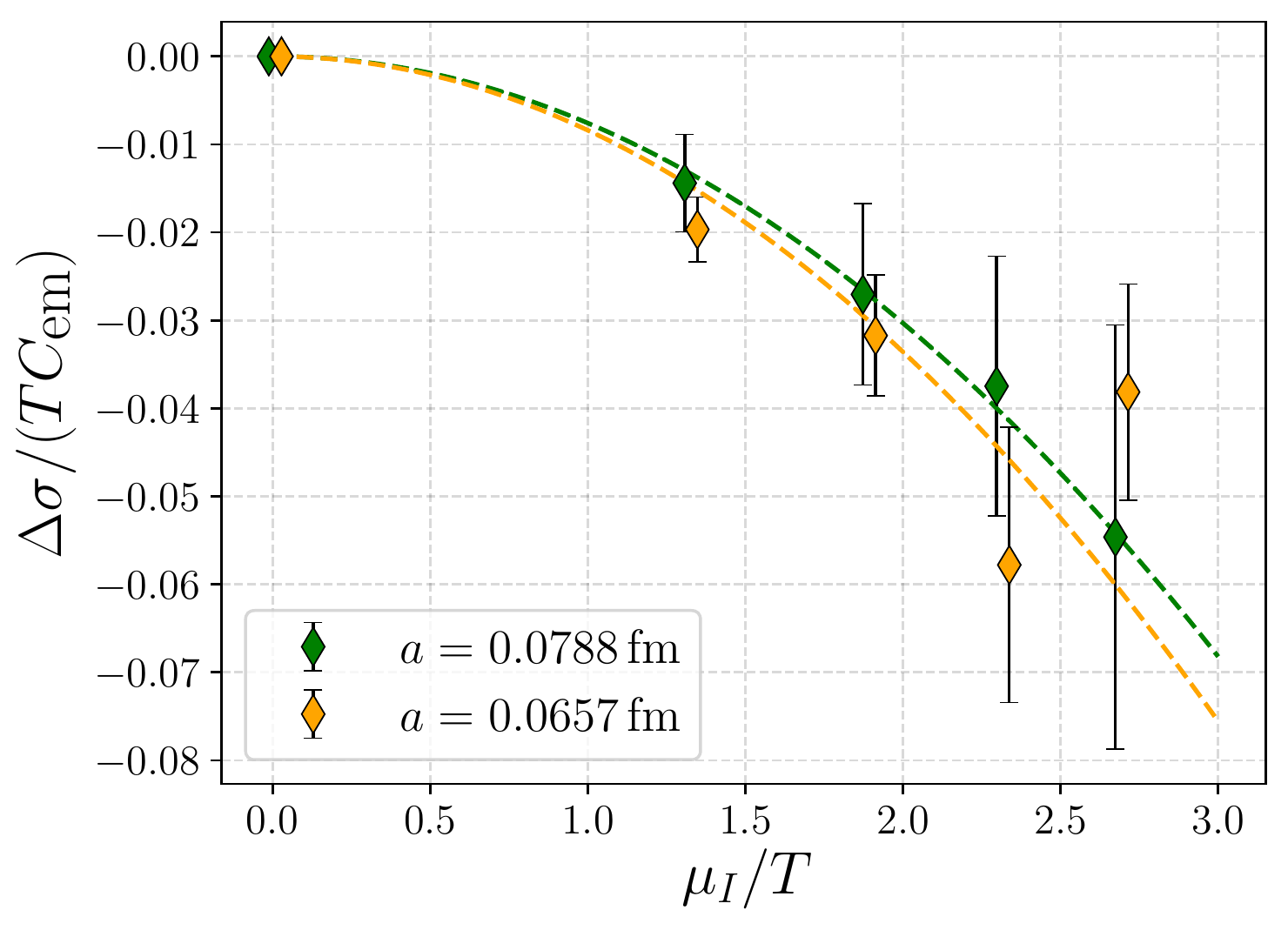}
    \caption{The electromagnetic conductivities due to non-zero imaginary baryon chemical potential $\Delta \sigma= \sigma_{\mu_I}-\sigma_{\mu_I=0}$ 
    as a function of $\mu_I/T$
    normalized to $T C_{em}$ for $T=200$\,MeV (left figure) and $T=250$\,MeV (right figure). In order to study the dependence of our results on lattice spacing, the calculations were carried out at two lattices $10\times 48^3$ and $12\times 48^3$ for each temperature.
    }
    \label{fig:cond}
\end{figure}

The change of the electromagnetic conductivity due to non-zero imaginary baryon chemical potential $\Delta \sigma= \sigma_{\mu_I}-\sigma_{\mu_I=0}$ normalized to $T C_{\mbox{\small em}}$ ($C_{\mbox{\small em}}=e^2\sum_f q_f^2$) at temperatures $T=200,\,250$\,MeV  is shown in Fig.~\ref{fig:cond}. 

Now few comments are in order. 
\begin{enumerate}
    \item In order to study discretization effects we carried out the study for two lattice spacings for each temperatures. For $T=200$\,MeV $a=0.0820, 0.0988$\,fm whereas and for $T=250$\,MeV $a=0.0657, 0.0788$\,fm (see lattice parameters in Table~\ref{tab:params}). From Fig.\,\ref{fig:cond} one sees that the results obtained at different lattice spacings agree within the uncertainties. 
    \item It is also seen that the results are well described by the quadratic polynomial
    \beq
     \frac {\Delta \sigma} {T C_{em}} = - c(T) \biggl ( \frac {\mu_I} {T} \biggr )^2,
     \label{delta_sigma}
    \eeq
    which after analytical continuation to real chemical potential becomes
    \beq
     \frac {\Delta \sigma} {T C_{em}} = c(T) \biggl ( \frac {\mu_B} {T} \biggr )^2.
     \label{delta_sigma1}
    \eeq
    Because of the uncertainties we don't see the dependence of the coefficient $c(T)$ on temperature (see Table~\ref{tab:params}).
    \item To overcome the sign problem our simulation are carried out at imaginary chemical potential. Notice also that for all lattice parameters under study, the coefficient $c(T)>0$. So, after analytical continuation one can conclude that the conductivity raises with baryon chemical potential. This conclusion is in agreement with the expectation that baryon chemical potential introduces additional fermion states to QGP which leads to rise of the conductivity. 
\end{enumerate}

At the end of this preprint let us discuss the values of the coefficient $c(T)$ which are presented in Table.\,\ref{tab:params}. If one takes into account that the characteristic value of the ratio $\sigma(T)/TC_{em}$ in the region $T \in (200-250)$~MeV is $\sim 0.2$ (see Fig.~3 in the review \cite{Aarts:2020dda}), the conductivity at $\mu_{u,d} \sim T$ raises by $\sim 30\%$ as compared to zero baryon density\footnote{Notice that in the report on the conference Lattice 2021 we incorrectly plotted the scale on the $\mu_I/T$ axis in Fig.\,\ref{fig:cond}. This led to incorrect values of the coefficients $c(T)$. }. We believe that this is quite considerable response of transport properties of QGP to non-zero baryon density. Our results for the $c(T)$ are in a reasonable agreement with the results of papers \cite{Steinert:2013fza, Srivastava:2015via} and they are larger than that in papers \cite{Soloveva:2019xph, Tripolt:2018jre}. We would like also to mention the paper \cite{Buividovich:2020dks} where the authors carried out lattice study of the conductivity in dense two-color QCD. Their results for the $c(T)$ are smaller than our values.

\section{Conclusion}

In this preprint we present our results on the study of the electromagnetic conductivity in dense quark-gluon plasma obtained within lattice simulations with $N_f = 2 + 1$ dynamical quarks. We employ stout improved rooted staggered quarks at the physical
point and the tree-level Symanzik improved gauge action.
The simulations are performed at imaginary chemical potential, and the
Tikhonov regularisation method is used to extract the conductivity from current-current correlators.
The results for the conductivity extracted in this way are analytically continued to real values of baryon chemical potential.
Our study indicates that 
electromagnetic conductivity of quark-gluon plasma raises with real baryon density, and this dependence is quite strong.

\begin{acknowledgments}
This work was supported by RFBR grant 18-02-40126.  Numerical simulations have been carried out on the MARCONI100 machine at CINECA, based on the agreement between INFN and CINECA (under project INF20\_npqcd and INF21\_npqcd). In addition we used computing resources of the Federal collective usage center Complex for Simulation and Data Processing for Mega-science Facilities at NRC ``Kurchatov Institute'', \url{http://ckp.nrcki.ru/}; 
the Supercomputer  ``Govorun'' of Joint Institute for Nuclear Research and the equipment of the shared research facilities of HPC computing resources at Lomonosov Moscow State University.
\end{acknowledgments}

\bibliographystyle{unsrt}
\bibliography{ref}

\end{document}